Original Article

# The cyclicality of loan loss provisions under three different accounting models: the United Kingdom, Spain, and Brazil

**Antônio Maria Henri Beyle de Araújo**
Universidade Católica de Brasília, Departamento de Ciências Contábeis, Brasília, DF, Brazil
Email: henri.beyle@uol.com.br

**Paulo Roberto Barbosa Lustosa**
Universidade de Brasília, Faculdade de Economia, Administração, Contabilidade e Gestão de Políticas Públicas, Departamento de Ciências Contábeis e Atuariais, Brasília, DF, Brazil
Email: lustosa@unb.br

**Edilson Paulo**
Universidade Federal da Paraíba, Centro de Ciências Sociais Aplicadas, Departamento de Finanças e Contabilidade, João Pessoa, PB, Brazil
Email: e.paulo@uol.com.br

## ABSTRACT

A controversy involving loan loss provisions in banks concerns their relationship with the business cycle. While international accounting standards for recognizing provisions (incurred loss model) would presumably be pro-cyclical, accentuating the effects of the current economic cycle, an alternative model, the expected loss model, has countercyclical characteristics, acting as a buffer against economic imbalances caused by expansionary or contractionary phases in the economy. In Brazil, a mixed accounting model exists, whose behavior is not known to be pro-cyclical or countercyclical. The aim of this research is to analyze the behavior of these accounting models in relation to the business cycle, using an econometric model consisting of financial and macroeconomic variables. The study allowed us to identify the impact of credit risk behavior, earnings management, capital management, Gross Domestic Product (GDP) behavior, and the behavior of the unemployment rate on provisions in countries that use different accounting models. Data from commercial banks in the United Kingdom (incurred loss), in Spain (expected loss), and in Brazil (mixed model) were used, covering the period from 2001 to 2012. Despite the accounting models of the three countries being formed by very different rules regarding possible effects on the business cycles, the results revealed a pro-cyclical behavior of provisions in each country, indicating that when GDP grows, provisions tend to fall and vice versa. The results also revealed other factors influencing the behavior of loan loss provisions, such as earning management.

**Correspondence address:**

**Antônio Maria Henri Beyle de Araújo**
Universidade Católica de Brasília, Departamento de Ciências Contábeis
SQSW 306, Bloco G, Apartamento 204 – CEP: 70673-437
Sudoeste – Brasília – DF – Brazil

# 1. INTRODUCTION

According to Longbrake and Rossi (2011), many of the rules that regulate the financial markets contributed to deepening the crisis that affected the global economy in the period between 2007 and 2009, fueling the debate regarding the way the financial system accentuates the effects of expansionary and contractionary phases in the economy. At the heart of the claim lies the concern regarding pro-cyclicality, which represents a positive correlation between a particular variable and economic activity (Bebczuk & Sangiácomo, 2010).

For Harrald and Sandal (2010), pro-cyclicality is the way that the banking system interacts with the real economy and accentuates the effects of an economic cycle. It derives from incentives to accelerate lending in periods of economic expansion and to reduce such activity in times of crisis.

According to Longbrake and Rossi (2011), the financial system should ideally dampen and not amplify business cycles. Pro-cyclicality becomes especially problematic when it accentuates the effect of a fall in the economic cycle and worsens a crisis. Thus, countercyclical rules, which reduce economic imbalances and the depth of business cycles, are welcome.

Also according to these authors, factors that explain the influence of the financial system on the economic environment include the accounting rules that set loan loss provisions. The expression "loan loss provision" ("*provisão para créditos de liquidação duvidosa*" in Portuguese) is the term used by the Brazilian Central Bank for the estimated losses from loans granted by financial institutions. In international accounting standards, this provision is represented by impairment losses derived from loans and receivables.

For Bikker and Metzemakers (2004), the provision leaves room for subjective judgment and allows for discretion when establishing levels considered as "adequate". The study by these authors identified a direct relationship between the provision set by banks and economic cycles, revealing that income smoothing, capital management, and tax rules also determine the level of provision.

Betancourt and Baril (2009) observe that at the beginning of a downward economic cycle there is little provision recognized in banks earnings. As the crisis worsens the provisions grow, deteriorating the equity situation of banks and reducing levels of lending when the market needs funds the most. This pro-cyclical behavior is adopted by banks that follow international accounting standards for provisions.

The accounting principles in the United States and those of the International Accounting Standards Board (IASB) adopt the incurred loss model, establishing that a loss in credit operations will only be recognized in earnings if its occurrence (related to a loss event) is likely and its value can be estimated with certainty.

According to Betancourt and Baril (2009), an alternative to the incurred loss model is the expected loss model, which recognizes provisions based on expected future losses, independently of objective evidence. This model uses as a reference the dynamic provision adopted by the Banco de España, which accepts the setting of a generic provision to protect banks against possible additional losses in a particular economic cycle.

In Brazil, Resolution n. 2,682/1999 of the National Monetary Council rules that the provision should be set based on the operation risk level and its value should be reviewed at least monthly, depending on the delay in the payment of principal or interest. This involves a mixed system with characteristics of expected loss (provisions based on future losses expectations) and incurred loss (provisions based on objective evidence represented by delays in payment).

The three accounting models would cause different behaviors of the provision in relation to economic cycles. In order to confirm this assumption, the relationship between the accounting criteria for setting provisions in banks and economic cycles should be investigated.

Studies such as those by Bikker and Metzemakers (2004) and by Boutavier and Lepetit (2007) have already been conducted to investigate the relationship between loan loss provisions and economic cycles. Besides being restricted to certain aspects of the provision, these studies used relatively different models and variables, thus leaving room for new analyses on the topic.

There are few studies involving the Brazilian model, which is presumably countercyclical since most of the value of the provisions is the result of classifying operations by level of risk at the time of lending and not migration between the risk categories due to credit defaults.

The aim is to analyze whether the accounting models involved in setting loan loss provisions in banks are pro-cyclical, countercyclical, or acyclical, using an econometric model with a consistent theoretical framework and empirical construction.

A panel data econometric model was specified in which the loan loss provision expense in relation to total assets is regressed against the variation in Gross Domestic Product (GDP), under the control of the unemployment rate and accounting variables that affect the provision. An identical specification is run for samples from three different provision accounting systems (United Kingdom, Spain, and Brazil) with the aim of verifying whether the covariance between the provision and GDP is significantly positive (countercyclical), negative (pro-cyclical), or without significance (acyclical).

The results indicated a negative and significant association between the provision expense and GDP in the three countries, signaling that the empirically confirmed pro-cyclical behavior is independent of the provision model adopted. One probable assumption regarding this finding is that the banks may be using discretion and engaging in earnings management via the provision and more than compensating for the expected effect of the models, but this is a hypothesis to be investigated in future research.

Besides this introduction, this paper presents the literature review in section 2; section 3 formulates the three research hypotheses; section 4 details the methodology; section 5 presents, analyzes, and discusses the results; and section 6 offers some final remarks.

## 2. CONSIDERATIONS REGARDING ECONOMIC CYCLES AND PRO-CYCLICALITY

A classic definition of "economic cycle", or "business cycle", is that of Burns and Mitchell (1946):

> Business cycles are a type of fluctuation found in the aggregate economic activity of nations that organize their work mainly in business enterprises: a cycle consists of expansions occurring at about the same time in many economic activities, followed by similarly general recessions, contractions, and revivals which merge into the expansion phase of the next cycle; in duration business cycles vary from more than one year to ten or twelve years.

For Harrald and Sandal (2010), pro-cyclicality is the way in which the banking system interacts with the real economy, accentuating the effects of an economic cycle. The pro-cyclicality cost depends on the extent to which the effects of the cycle are accentuated and the fact that these effects are damaging to the economy. Pro-cyclicality derives from incentives for the acceleration of lending in periods of expansion and the reduction in such activity in crises.

According to Bebczuk et al. (2010), pro-cyclicality is a synchronized movement in the growth of credit and production, both in time $t$. Excessive cyclicality in lending can lead to problems such as exacerbating the economic cycle, an increase in systemic risk, and the inadequate allocation of resources for loans.

For Gonzales (2009), pro-cyclicality is present when the actions of banks tend to reinforce the movements of the underlying business cycles, indicating a correlation between the aforementioned actions and the current economic environment. Pro-cyclical measures are, therefore, those that contribute to strengthening the current economic cycle.

## 3. LOAN LOSS PROVISIONS IN BANKS

Hendriksen and Van Breda (1999) state that an important element in evaluating receivables is the treatment of uncertainty with regards to their payment by the client. Thus, revenue should be measured by the value that it is expected will be received. In the quality of the value reserve, the provision adjusts the gross value of loans according to their credit risk. This adjustment has a direct effect on the income calculation, affecting the amount to be paid out to shareholders.

### 3.1. Provision Models

Banks' provision rules can be: (i) backward-looking, based on losses incurred in operations and considering the events and conditions prior to the balance sheet; and (ii) forward-looking, set by considering expected losses and based on trend analyses (Bouvatier & Lepetit, 2012).

According to Cortavarria, Dziobek, Kanaya, and Song (2000), provisions can be generic and specific. Generic

provisions are possible or latent losses not yet identified (*ex-ante*) and applicable to all operations. Specific provisions reflect individually identified losses related to an observable event (*ex-post*).

## 3.2. Earnings Management via Provisions

Accounting rules allow for the use of discretion and judgment in the financial statements preparation (Cohen & Zarowin, 2007). This enables the use of business knowledge and expertise to choose the most appropriate accounting methods, estimates, and disclosure. The problem is when managers find incentives to convey information to their own benefit, choosing accounting methods and accounting estimates that do not adequately reflect the underlying economic conditions.

For Cohen and Zarowin (2007), this accounting information discretion can be both a value maximizer and opportunistic, leaving room for "earnings management". Earnings management can be achieved via the manipulation of accruals, which according to Martinez (2001, p. 29) represent "the difference between net income and net cash flow", or via real transactions, such as a reduction in spending on research and development. In the case of provision, earnings management occurs through accruals manipulation. For Martinez (2001), when it is left to managers to make accounting choices, they may not limit themselves to accounting facts alone and use exogenous variables, which will also influence the decision. Thus, provision will also consider aspects that extrapolate credit quality.

Studies have revealed indications of earnings management being practiced by financial institutions in Brazil, based on the use of loan loss provisions. Macedo and Kelly (2016) identified that from 2006 to 2012 the income level obtained by financial institutions influenced these provisions. Bortoluzzo, Sheng, and Gomes (2016) found similar results by analyzing the behavior of 123 Brazilian banks between 2001 and 2012. Silva (2016) showed that between June 2009 and December 2014 Brazilian and Luso-Spanish banking institutions used provision expenses to manage earnings. Cursio and Hasan (2015) found that from 1996 to 2006 earnings management was shown to be an important determinant of the provisions of financial intermediaries that operate in the Euro zone.

## 3.3. Capital Management via Provisions

In the capital management hypothesis, banks set higher provisions when their regulatory capital is relatively low. According to Bikker and Metzemakers (2004), this occurs because the Basel Accord allows for the accumulated provisions in the balance sheet, although limited to a percentage of risk-weighted assets, to comprise banks' regulatory capital. This allowance occurs only in relation to level 2 capital, which is the lowest quality regulatory capital and whose value cannot exceed the value of level 1 capital, taken as main capital.

## 3.4. Studies Relating the Provision to Economic Cycles

Table 1 lists studies involving the pro-cyclicality of the provision in banks.

**Table 1** *Summary of the main articles involving the analysis of the pro-cyclicality of the loan loss provision in financial institutions.*

| Author/Period | Methodology | Results |
|---|---|---|
| Cavallo and Majnoni<br>Period: 1988 to 1999<br>Publication: 2001 | Estimation of an equation in which the provision is a function of: (i) specific variables of the banks, (ii) macroeconomic variables, and (iii) institutional variables, with the use of three different techniques (OLS cross section, pooled cross section, and panel fixed effect estimation). | Strong evidence that the relationship between the provision and the earnings of banks from the G-10 countries presents a positive sign. The same sign is negative for banks from countries outside the G-10, which on average present much smaller provisions in good periods and are forced to increase them in bad periods. |
| Bikker and Hu<br>Period: 1979 to 1999<br>Publication: 2002 | To identify the relationship between the provision and the business cycle, three macroeconomic variables (real GPD growth, unemployment, and inflation) and three specific variables of the banking sector (loans, net financial intermediation earnings, and defaults) were used as explanatory variables for the provision in the linear regression. | Increases in the provision depend a lot on the business cycle. In bad times, the provisions increase. Banks contribute significantly more to the provisions in years of relatively higher earnings (as a precaution or as a way of smoothing income), leading the sector to be less pro-cyclical than it should. |
| Laeven and Majnoni<br>Período: 1988 to 1999<br>Publication: 2003 | In the linear regression, the provisions are a result of the earnings before income tax and of the provisions for losses; of the real growth in loans; of the real growth in GDP *per capita*; and of the year dummies. | Many banks delay the recognition of provisions as much as possible, recording them when an economic contraction has just taken hold and thus increasing the impact of economic cycles on earnings and on capital. |
| Bikker and Metzemakers<br>Period: 1991 to 2001<br>Publication: 2004 | Based on the models by Cavallo and Majnoni (2002) and by Laeven and Majnoni (2003), the authors worked with additional variables such as GDP growth and dummies for the countries. The use of the provision for capital management was also tested. | The provisions of banks are usually substantially greater in periods of lower GDP growth, reflecting the growing risk of their credit portfolios when the economic cycle changes and enters into a downturn. This effect is mitigated by an increase in the provision in periods of higher earnings and in those in which the growth in the volume of lending becomes greater. |
| Handorf and Zhu<br>Period: 1990 to 2000<br>Publication: 2006 | The model is based on the assumption that the provision (dependent variable) is a linear function of two variables: (i) the initial value of the accumulated provision, net of the falls or losses *ex-post* of the current period; and of management's expectations in relation to future falls, based on current available information. | There is a positive correlation between the provision and GDP. The empirical tests do not support the pro-cyclicality of the provision in banks. The results were different depending on the size of the institutions. Medium-sized banks tend to use information on projected losses when defining the value of the provision (countercyclical attitude). Smaller banks and much bigger banks tend towards provision practices that consider current losses (pro-cyclical attitude). |
| Bouvatier and Lepetit<br>Period: 1992 to 2004<br>Publication: 2007 | The model evaluates whether the evolution of the provision explains the changes in banks' behavior in relation to lending during the economic cycle. The linear regression model estimated the discretionary and non-discretionary components of the provision. The authors introduced the lagged dependent variable as an explanatory variable, in order to consider a dynamic adjustment in the provision. Capital management, earnings management, and the signaling of equity robustness are also important variables. | Setting provisions to cover expected future losses in loans ("non-discretionary provisions") causes major fluctuations in credit. The non-discretionary component of the provision amplifies the credit cycle: in an economy rising phase, banks tend to underestimate credit risks, indicating the existence of greater incentives for the granting of new loans since the costs of lending are undervalued. Provisions with managerial purposes (discretionary provisions) do not produce the same effect. |
| Glen and Mondrágon-Vélez<br>Period: 1996 and 2008<br>Publication: 2011 | The effects of business cycles on the performance of banks' loans portfolios in developing countries were analyzed using linear and non-linear models. The provision was the proxy for the portfolio's performance. GDP and other macroeconomic variables were considered as explanatory variables. The interest rate on loans and a set of variables with the individual characteristics of the banks of each country were also considered. | While economic growth is the main driver in the performance of banks' credit portfolios, the effects caused by the interest rates are second order. The relationship between the provisions and economic growth is highly linear only in economic conditions of extreme stress. Greater loss provisions are related to the private sector leverage level, to the bad quality of a loans portfolio, and to the absence of penetration and capitalization of the financial system. |

GDP = gross domestic product.
**Source:** *Elaborated by the authors.*

# 4. RESEARCH HYPOTHESES

The hypotheses to be tested in this study and formulated in an alternative form are:

**H₁:** Provisions based on the incurred loss model present a negative relationship with the variables that represent economic activity, contributing towards enhancing possible contractions experienced by the economy, or towards amplifying the effects of favorable economic cycles. They are pro-cyclical in nature.

**H₂:** Provisions based on the expected loss model present a positive relationship with the variables that represent economic activity, preventing possible contractions in the economy from being enhanced, or the effects of favorable economic cycles from being amplified. They are countercyclical in nature.

**H₃:** Provisions based on the mixed model adopted in Brazil present a positive relationship with the variables that represent economic activity, preventing possible contractions experienced by the economy from being enhanced, or the effects of favorable economic cycles from being amplified. They are countercyclical in nature.

# 5. METHODOLOGY

## 5.1. Presentation of the Model and Discussion of the Variables

A linear regression model was defined to evaluate the relationship between the variables of interest in order to identify how the provisions are associated with the variables that characterize business cycles. Most of the studies relating provisions and economic cycles use this type of model, such as Cavallo and Majnoni (2001), Bikker and Hu (2002), Laeven and Majnoni (2003), Bikker and Metzemakers (2004), Handorf and Zhu (2006), Bouvatier and Lepetit (2007), and Glen and Mondragón-Vélez (2011).

The model is estimated using the panel data method. The empirical tests use the model below, elaborated based on the aims and hypotheses of the research:

$$PROV_{it} = \beta_0 + \beta_1 \Delta GDP_t + \beta_2 UNEMP_t + \beta_3 EAR_{it}$$

$$+ \beta_4 \Delta lnLOAN_{it} + \beta_5 LOAN_{it} + \beta_6 NE_{it} + \beta_7 SIZE_{it_i} + \varepsilon_{it}$$

$$\boxed{1}$$

where:
- $PROV_{it}$ = loan loss provision over banks' average total assets $i$ in time $t$.
- $\Delta GDP_t$ = real variation in GDP in time $t$.
- $UNEMP_t$ = unemployment rate in time $t$.
- $EAR_{it}$ = earnings before income tax, profit sharing, and loan loss provisions over banks' average total assets $i$ in time $t$.
- $\Delta lnLOAN_{it}$ = variation in the loan balances of banks $i$ in time $t$.
- $LOAN_{it}$ = balance of credit operations over total assets of banks $i$ in time $t$.
- $NE_{it}$ = equity over total assets of banks $i$ in time $t$.
- $SIZE_{it}$ = size of banks $i$ in time $t$.

The aim of the model is not to capture a possible cause and effect relationship between dependent variables and independent variables, but rather, to identify how the dependent variable behaves in relation to variations in each independent variable, especially the macroeconomic ones.

Despite the accounting variables being endogenous, their inclusion in the model is warranted because the provision is influenced by prominently accounting factors, such as receivables, earnings, and equity. Moreover, these variables work as control variables, helping to more adequately capture the relationship between the dependent and the independent variables, taken as critical to the study (GDP and unemployment rate).

Most of the models that study the provision-economic cycle relationship use the provision as a dependent variable. In this study, it represents the net amount of provision expenses set in the period, obtained by the difference between the set and the reverted provisions.

The independent variables can be classified into: (i) macroeconomic variables for the countries (real GDP growth and unemployment rate); (ii) accounting variables for the banks (earnings before income tax, profit sharing, and loan loss provisions over average total assets, variation in loan balances, loan balances over total assets, and equity over total assets); and (iii) a control variable (bank size).

The real variation in GDP is a critical variable in the model and considered a more useful indicator for representing the business cycle. Its value represents the percentage variation in GDP at constant prices.

The unemployment rate is another representative metric of the current phase of the economic cycle. In Brazil, it indicates the joblessness rate for 10-year-olds upwards, considering only the main metropolitan regions of the country.

The earnings over total assets variable, which indicates a possible use of the provision for earnings management, is the result of dividing the value of earnings before income tax, profit sharing, and loan loss provisions by the average value of total assets.

The growth in lending variable, which indicates the evolution of the banks' credit risk, represents the variation in the balances of credit operations in real terms. In the Brazilian case, the deflator used was the General Market Price Index (IGPM). For Spain and the United Kingdom, the Consumer Price Index (CPI) was used. The variation in the balances of loans was calculated by the difference in the natural logarithms in the following way: $\ln(\text{credit operations}_{it}/\text{IGPM}_{it}) - \ln(\text{credit operations}_{it-1}/\text{IGPM}_{it-1})$.

The loans over total assets variable considers the exposure of the banks to credit risks, indicating the relative size of the loans portfolio. It represents the division of the balance of receivables from credit operation accounts by the value of total assets.

The equity over total assets variable, inserted into the model to indicate a possible use of the provision as a capital management tool, is the result of dividing net equity by total assets.

The $SIZE_{it}$ variable, defined as the natural logarithm of total assets, deflated by the IGPM or the CPI, aims to control the size effects of the institutions. Larger banks, or those participating in conglomerates, are expected to set more robust provisions.

## 5.2. Summary of the Expected Results

Table 2 summarizes the expected results in relation to the behavior and to the sign of the coefficients of the explanatory variables, based on the formulated hypotheses.

**Table 2** *Summary of the expected results in relation to the model explanatory variables.*

| Variable | Expected Behavior | Sign |
|---|---|---|
| GDP Growth | In countries with provision accounting procedures that are considered pro-cyclical, such as the United Kingdom, a positive variation in GDP should make the level of the banks' provision decrease. | - |
| | In countries with provision accounting criteria that are considered countercyclical, such as Spain, a positive variation in GDP should make the level of the banks' provision increase. | + |
| | In Brazil, a country with provision accounting criteria that is considered mixed, a positive variation in GDP should make the level of the banks' provision increase, considering the prevalence of *ex-post* criteria over *ex-ante* criteria, which presupposes a prevalently countercyclical nature. | + |
| Unemployment rate | In countries with pro-cyclical provision accounting criteria such as the United Kingdom, a positive variation in unemployment should make the level of the banks' provision increase. | + |
| | In countries with countercyclical provision accounting criteria, such as Spain, a positive variation in unemployment should make the level of banks' provision decrease. | - |
| | In Brazil, a country with provision accounting criteria that is considered mixed, a positive variation in the unemployment rate should make the level of the banks' provision decrease, given the prevalence of *ex-post* criteria over *ex-ante* criteria, which presupposes a prevalently countercyclical nature. | - |

**Table 2** *Cont.*

| Variable | Expected Behavior | Sign |
|---|---|---|
| Earnings/Assets | Considering that banks use the provision as an earnings management tool, a positive variation in earnings before income tax, profit sharing, and loan loss provisions should make the level of the bank's provision increase. | + |
| NE | Considering that banks use the provision as a capital management tool, a negative variation in net equity should make the level of the banks' provision increase. A bank would be likely to record a higher provision was below that required. | - |
| Loans/Assets | Considering that a greater share of credit operations in total investments represents an increase in credit risk, a positive variation in the relationship between loans and total assets should make the level of commercial banks' provisions increase. | + |
| Variation in loans | Considering that a positive variation in the volume of credit operations over time represents an increase in credit risk, a positive variation in the volume of credit operations should make the level of commercial banks' provisions increase. | + |

*NE = net equity; GDP = gross domestic product.*
**Source:** *Elaborated by the authors.*

## 5.3. Tests Applied and Robustness Procedures Adopted

The Im, Pesaran, and Shin – I.P.S, ADF-Fisher, and PP-Fisher unitary root tests indicated that in the countries studied the risk of a spurious regression was removed.

The Pearson correlation matrix (tables 3, 4, and 5) proved that a high correlation (above 0.8) did not exist between the independent variables in the countries analyzed.

**Table 3** *Pearson Correlation Matrix – Brazil.*

| | LOAN | ΔlnLOAN | EAR | NE | ΔGDP | UNEMP | SIZE |
|---|---|---|---|---|---|---|---|
| LOAN | 1.000000 | - | - | - | - | - | - |
| ΔlnLOAN | 0.095587 | 1.000000 | - | - | - | - | - |
| EAR | 0.233714 | 0.024260 | 1.000000 | - | - | - | - |
| NE | -0.147020 | -0.051504 | 0.139431 | 1.000000 | - | - | - |
| ΔGDP | -0.011816 | 0.007476 | 0.025810 | 0.017003 | 1.000000 | - | - |
| UNEMP | -0.057126 | 0.039506 | 0.120247 | 0.045110 | 0.040903 | 1.000000 | - |
| SIZE | -0.097314 | 0.005905 | -0.107682 | -0.592854 | -0.009121 | -0.234355 | 1.000000 |

*LOAN = share of credit operations in the total assets of Brazilian banks; ΔlnLOAN = variation in the balances of credit operations of banks i in time t; EAR = share of earnings before income tax, profit sharing, and loan loss provisions over average total assets of the Brazilian banks; NE = equity over total assets of banks i in time t; ΔGDP = real growth in gross domestic product in time t; UNEMP = unemployment rate in time t; SIZE = size of the commercial bank, represented by the natural logarithm of total assets deflated by the general market price index.*
**Source:** *Elaborated by the authors.*

**Table 4** *Pearson correlation matrix – Spain.*

| | LOAN | ΔlnLOAN | EAR | NE | ΔGDP | UNEMP | SIZE |
|---|---|---|---|---|---|---|---|
| LOAN | 1.000000 | - | - | - | - | - | - |
| ΔlnLOAN | 0.179954 | 1.000000 | - | - | - | - | - |
| EAR | -0.088175 | 0.062070 | 1.000000 | - | - | - | - |
| NE | -0.113306 | -0.026035 | 0.241673 | 1.000000 | - | - | - |
| ΔGDP | 0.121625 | 0.119785 | 0.082047 | 0.004057 | 1.000000 | - | - |
| UNEMP | -0.191935 | -0.091091 | -0.126566 | 0.003399 | -0.739170 | 1.000000 | - |
| SIZE | -0.062023 | 0.011395 | -0.070449 | -0.646019 | -0.070423 | 0.059811 | 1.000000 |

*LOAN = share of credit operations in the total assets of Brazilian banks; ΔlnLOAN = variation in the balances of credit operations of banks i in time t; EAR = share of earnings before income tax, profit sharing, and loan loss provisions over average total assets of the Brazilian banks; NE = equity over total assets of banks i in time t; ΔGDP = real growth in gross domestic product in time t; UNEMP = unemployment rate in time t; SIZE = size of the commercial bank, represented by the natural logarithm of total assets deflated by the general market price index.*
**Source:** *Elaborated by the authors.*

**Table 5** *Pearson correlation matrix – United Kingdom.*

| | LOAN | ΔlnLOAN | EAR | NE | ΔGDP | UNEMP | SIZE |
|---|---|---|---|---|---|---|---|
| LOAN | 1.000000 | - | - | - | - | - | - |
| ΔlnLOAN | 0.136788 | 1.000000 | - | - | - | - | - |
| EAR | 0.053496 | 0.026894 | 1.000000 | - | - | - | - |
| NE | 0.114561 | -0.055089 | 0.015586 | 1.000000 | - | - | - |
| ΔGDP | -0.036011 | 0.110507 | -0.035462 | -0.003076 | 1.000000 | - | - |
| UNEMP | 0.012340 | -0.155514 | -0.030084 | 0.050401 | -0.607156 | 1.000000 | - |
| SIZE | -0.279305 | 0.042143 | -0.041180 | -0.448594 | -0.060661 | 0.097252 | 1.000000 |

*LOAN/ASS = share of credit operations in the total assets of Brazilian banks; ΔlnLOAN = variation in the balances of credit operations of banks i in time t; EAR = share of earnings before income tax, profit sharing, and loan loss provisions over average total assets of the Brazilian banks; NE = equity over total assets of banks i in time t; ΔGDP = real growth in gross domestic product in time t; UNEMP = unemployment rate in time t; SIZE = size of the commercial bank, represented by the natural logarithm of total assets deflated by the general market price index.*
**Source:** *Elaborated by the authors.*

The risk of multicollinearity was analyzed by applying the variance inflation test among the explanatory variables, which did not reveal multicollinearity problems.

The Chow test was carried out to test the existence of individual heterogeneity and confirm whether the use of panel data would apply to the study. In the three countries, the option was to use the regression with individual effects. Then the Hausman test was carried out to define the best panel data method for the regression estimation. For Brazil, the result indicated the random effects model. For Spain and the United Kingdom, it was the fixed effects model.

In order to analyze the existence of autocorrelation between the residuals of the regression, the Durbin-Watson test was used. In the three countries, the value of the statistic was between $d_l$ and $d_u$, indicating that the test was inconclusive. Adopting a conservative position, the null hypothesis of the inexistence of autocorrelation is rejected; that is, the model residuals appear to be autocorrelated.

In light of the possibility of the existence of cross-sectional autocorrelation of the residuals, the use of the SUR cross-sectional standard errors method (PCSE) in the model estimation arose as an alternative to the problem, enabling the generation of robust parameters even in the presence of residues autocorrelation.

The overall significance of the model was proven using the F test.

## 5.4. Sample Definition and Description of the Data Source

### 5.4.1. In relation to the banks that operate in Brazil.

In Brazil, the sample considered the institutions that formed part of Consolidated Banking I, formed by Banking Conglomerate I (composed of at least one institution of the Commercial Bank or Multiple Bank with Commercial Portfolio type) and Independent Banking Institutions I (Commercial Banks, Multiple Banks with Commercial Portfolio, and Savings Banks that do not form part of a conglomerate).

Ninety-eight institutions were initially considered. Together, these institutions controlled 84.1% of the total assets of the National Financial System on December 31st 2012.

Banco Plural, Morgan Stanley, Banco BM&F, Banco Opportunity, BNY Mellon, Western Union, and Banco Petra were excluded from the sample as they did not present a credit operations balance in the period analyzed.

Data was extracted from the semi-annual financial statements of the chosen institutions on June 30st and December 31st between 2001 and 2012, obtained from the Banco Central do Brasil website.

### 5.4.2. In relation to the banks that operate in Spain.

The sample considered all of the Spanish banks associated with the Asociación Española de Banca (AEB). Fifty-eight institutions, including financial conglomerates and individuals banks, were considered in the study.

### 5.4.3. In relation to the banks that operate in the United Kingdom.

The selection of banks that operate in the United Kingdom was based on the publication "*List of Banks as Compiled by the Bank of England on 31 March 2013*", which covers the list of banks under the supervision of the Bank of England, available from the website www.bankofengland.co.uk. The list is comprised of 153 financial institutions, with 45 banks being selected to compose the sample (29.41% of the supervised institutions). The sample prioritized institutions closing their financial year on December 31st and which presented their information in identical monetary bases (pounds sterling). The information was obtained from the Company Check website (www.companycheck.co.uk), which provides the annual financial statements of banks operating in the United Kingdom.

# 6. RESULTS

## 6.1. Descriptive Statistics of the Dependent Variable

The descriptive statistics of the dependent variable are found in Table 6. From the second semester of 2001 to the second semester of 2012, the commercial banks that operated in Brazil, in Spain, and in the United Kingdom set, on average, provisions in percentages equal to 0.83%, 0.2668%, and 0.3607% of total assets, respectively. The standard deviations show a large variability in the provision, perhaps caused by the reversion mechanism, and indicate that the frequency distributions of the provision in the three countries have a similar behavior.

**Table 6** *Descriptive statistics of the dependent variable in the model – Brazil, Spain, and the United Kingdom, from 2001 to 2012.*

| Provisions/Total Assets | Brazil | Spain | United Kingdom |
|---|---|---|---|
| Mean | 0.0083 | 0.0027 | 0.0036 |
| Median | 0.0044 | 0.0008 | 0.0013 |
| Maximum Value | 0.1643 | 0.0858 | 0.1432 |
| Minimum Value | -0.0519 | -0.0337 | -0.0407 |
| Standard Deviation | 0.0146 | 0.0071 | 0.0102 |

**Source:** *Elaborated by the authors.*

## 6.2. Descriptive Statistics of the Independent Variables

### 6.2.1. Brazil.

According to Table 7, on average only 34.34% of the resources were used for credit operations, a lower percentage than those of the banks that operate in Spain and the United Kingdom, which were around 82% and 71%, respectively.

The average share of own resources over the total investments of the commercial banks has shown to be quite significant, at around 23%, and more comfortable than that of the banks that operate in Spain (average of 20%) and much higher than that of the banks that operate in the United Kingdom (average of 13.8%).

**Table 7** *Descriptive statistics of the independent variables in the model – Brazil, 2001-2012.*

| MEASURE | LOAN | ΔlnLOAN | NE | EAR | ΔGDP | UNEMP | SIZE |
|---|---|---|---|---|---|---|---|
| Mean | 0.3434 | 0.1272 | 0.2320 | 0.0225 | 0.0167 | 0.0855 | 13.999 |
| Median | 0.3309 | 0.1268 | 0.1551 | 0.0184 | 0.0196 | 0.0830 | 13.922 |
| Maximum Value | 1.0344 | 5.6920 | 0.9999 | 0.3568 | 0.0474 | 0.1300 | 20.8069 |
| Minimum Value | 0.0000 | -5.2298 | -0.1038 | -0.1831 | -0.0279 | 0.0470 | 8.2451 |
| Standard Deviation | 0.2356 | 0.5020 | 0.2154 | 0.0326 | 0.0174 | 0.0222 | 2.3498 |

*LOAN = share of credit operations in the total assets of the Brazilian banks; ΔlnLOAN = real growth in the balances of credit operations recorded in the balance sheets of the Brazilian banks; NE = share of equity over total assets of the Brazilian banks; EAR = share of earnings before income tax, profit sharing, and loan loss provisions over the average total assets of the Brazilian banks; ΔGDP = real growth in the gross domestic product of Brazil; UNEMP = unemployment rate in Brazil; SIZE = size of the commercial bank, represented by the natural logarithm of total assets deflated by the general market prices index.*

**Source:** *Elaborated by the authors.*

### 6.2.2. Spain.

Table 8 presents the descriptive statistics of the independent variables in relation to the banks that operate in Spain. The average return on assets was around 0.53%, below the returns of the banks that operate in Brazil (2.25%) and in the United Kingdom (1.91%).

**Table 8** *Descriptive statistics of the independent variables in the model – Spain, 2001-2012.*

| MEASURE | LOAN | ΔlnLOAN | NE | EAR | ΔGDP | UNEMP | SIZE |
|---|---|---|---|---|---|---|---|
| Mean | 0.8195 | 0.0534 | 0.1998 | 0.0053 | 0.0078 | 0.1362 | 14.0939 |
| Median | 0.8793 | 0.0354 | 0.0808 | 0.0036 | 0.0140 | 0.1130 | 13.9648 |
| Maximum Value | 1.0766 | 8.0520 | 0.9999 | 0.3045 | 0.0201 | 0.2510 | 20.9620 |
| Minimum Value | 0.0000 | -8.0301 | -0.0315 | -0.1705 | -0.0278 | 0.0810 | 0.0810 |
| Standard Deviation | 0.1905 | 0.4701 | 0.2745 | 0.0242 | 0.0131 | 0.0527 | 2.4123 |

*LOAN = share of credit operations in the total assets of the Spanish banks; ΔlnLOAN = real growth in the balances of credit operations recorded in the balance sheets of the Spanish banks; NE = share of equity over total assets of the Spanish banks; EAR = share of earnings before income tax, profit sharing, and loan loss provision over the average total assets of the Spanish banks; ΔGDP = real growth in the gross domestic product of Spain; UNEMP = unemployment rate in Spain; SIZE = size of the commercial bank, represented by the natural logarithm of total assets deflated by the general market price index.*
**Source:** *Elaborated by the authors.*

### 6.2.3. United Kingdom.

The descriptive statistics of the independent variables related to the banks that operate in the United Kingdom are presented in Table 9.

**Table 9** *Descriptive statistics of the independent variables in the model – United Kingdom, 2001-2012.*

| MEASURE | LOAN | ΔlnLOAN | NE | EAR | ΔGDP | UNEMP | SIZE |
|---|---|---|---|---|---|---|---|
| Mean | 0.7076 | 0.1129 | 0.1380 | 0.0191 | 0.0150 | 0.0600 | 13.2893 |
| Median | 0.7453 | 0.0900 | 0.0935 | 0.0086 | 0.0224 | 0.0533 | 12.8845 |
| Maximum Value | 1.3041 | 6.5087 | 0.9885 | 5.5899 | 0.0395 | 0.0807 | 20.6667 |
| Minimum Value | 0.0000 | -2.0549 | 0.0000 | -0.1625 | -0.0517 | 0.0467 | 3.98689 |
| Standard Deviation | 0.2353 | 0.4501 | 0.1439 | 0.2472 | 0.0242 | 0.0140 | 2.65957 |

*LOAN = share of credit operations in the total assets of the British banks; ΔlnLOAN = real growth in the balances of credit operations recorded in the balance sheets of the British banks; NE = share of equity over total assets of the British banks; EAR = share of earnings before income tax, profit sharing, and loan loss provisions over the average total assets of the British banks; ΔGDP = real growth in the gross domestic product of the United Kingdom; UNEMP = unemployment rate in the United Kingdom; SIZE = size of the commercial bank, represented by the natural logarithm of total assets deflated by the general market price index.*
**Source:** *Elaborated by the authors.*

## 6.3. Test of the Hypotheses

### 6.3.1. H1: Banks that operate in the United Kingdom

The coefficient of determination ($R^2$) indicates that the independent variables are associated with 38.87% of the behavior of the dependent variable. The F statistic, with a p-value of 0.0000, confirms the statistical significance of the model (Table 10).

**Table 10** *Coefficients of Determination and F Statistic of the model – United Kingdom.*

| Variables | Value |
|---|---|
| $R^2$ | 0.3887 |
| $R^2$ Adjusted | 0.3191 |
| F Statistic | 5.5863 |
| P-Value (F) | 0.0000 |

**Source:** *Elaborated by the authors.*

The macroeconomic variables and the specific accounting variables of the banks, except GDP, do not have significant effects on the provision in the United Kingdom. The "loans over total assets" and "growth in lending" variables were not found to be significant, thus going against the expectations presented in Table 2.

The "earnings before income tax, profit sharing, and loan loss provisions over total assets" variable was also not found to be significant, indicating that variations in earnings would not explain the behavior of the provision and going against the expectations of the research. Earnings management via provisions does not appear to be a common practice among banks in the United Kingdom.

**Table 11** *Regression results – United Kingdom.*

| Variable | Coefficient | Standard deviation | T statistic | Prob. |
|---|---|---|---|---|
| C | 0.0114 | 0.0074 | 1.5429 | 0.1236 |
| LOAN | -0.0009 | 0.0023 | -0.3715 | 0.7105 |
| ΔlnLOAN | -0.0006 | 0.0019 | -0.3013 | 0.7633 |
| EAR | -0.0022 | 0.0053 | -0.4124 | 0.6803 |
| NE | -0.0052 | 0.0072 | -0.7189 | 0.4726 |
| ΔGDP | -0.0553 | 0.0140 | -3.9372 | 0.0001 |
| UNEMP | 0.0505 | 0.0270 | 1.8662 | 0.0627 |
| SIZE | -0.0007 | 0.0005 | -1.2307 | 0.2191 |

*C = constant in the regression; LOAN = balances of credit operations of banks i in time t; ΔlnLOAN = variation in the balances of credit operations of banks i in time t; EAR = earnings before income tax, profit sharing, and loan loss provisions over the average total assets of banks i in time t; NE = equity over the total assets of banks i in time t; ΔGDP = real growth of gross domestic product in time t; UNEMP = unemployment rate in time t; SIZE = size of banks i in time t.*
**Source:** *Elaborated by the authors.*

The "net equity over total assets" variable was not found to be significant, thus going against the capital management hypothesis.

At a 1% level of significance, the coefficient of the "GDP growth" variable was found to be significantly negative, indicating pro-cyclical behavior of the banks in the United Kingdom, in line with hypothesis H1.

The "unemployment rate growth" variable was only found to be significant at a 10% level of significance, thus going against the expectations. This result, however, supports the findings of previous studies, such as Bikkers and Metzemakers (2004).

The "size" control variable was also not found to be significant.

### 6.3.2. H2: Banks that operate in Spain.

With regards to the banks that operate in Spain, the coefficient of determination ($R^2$) indicates that the independent variables are associated with 38.31% of the dependent variable behavior. The F statistic, with a p-value of 0.0000, confirms the statistical significance of the model (Table 12).

**Table 12** *Coefficients of determination and F statistic of the model – Spain.*

| Variables | Value |
|---|---|
| $R^2$ | 0.3831 |
| $R^2$ Adjusted | 0.3436 |
| F Statistic | 9.7147 |
| P-Value (F) | 0.0000 |

**Source:** *Elaborated by the authors.*

The "loans over total assets", "earnings over total assets", "GDP growth" and "size" variables have significant effects on loan loss provisions in Spain, at different levels of significance, as can be observed in Table 13. The other variables were not significant.

**Table 13** *Regression results – Spain.*

| Variable | Coefficient | Standard deviation | t-statistic | Prob. |
|----------|-------------|--------------------|-------------|-------|
| C | -0.0149 | 0.0074 | -2.0191 | 0.0438 |
| LOAN | 0.0031 | 0.0017 | 1.8418 | 0.0658 |
| ΔlnLOAN | -0.0004 | 0.0003 | -1.2059 | 0.2281 |
| EAR | -0.1004 | 0.0170 | -5.9089 | 0.0000 |
| NE | -0.0020 | 0.0052 | -0.3825 | 0.7022 |
| ΔGDP | -0.0534 | 0.0192 | -2.7727 | 0.0057 |
| UNEMP | 0.0040 | 0.0049 | 0.8224 | 0.4111 |
| SIZE | 0.0011 | 0.0004 | 2.5368 | 0.0113 |

**C** = constant in the regression; *LOAN* = balances of credit operations of banks i in time t; *ΔlnLOAN* = variation in the balances of credit operations of banks i in time t; *EAR* = earnings before income tax, profit sharing, and loan loss provisions over the average total assets of banks i in time t; *NE* = equity over the total assets of banks i in time t; *ΔGDP* = real growth of gross domestic product in time t; *UNEMP* = unemployment rate in time t; *SIZE* = size of banks i in time t.
**Source:** *Elaborated by the authors.*

The "loans over total assets" variable was not found to be significant, at a 10% level of significance. The positive sign of the coefficient supports the expectation presented in Table 2, indicating that the provision tends to rise when the share of loans over total assets increases.

Contrary to expectations, the "growth in lending" variable was not found to be significant, indicating that the variation in the volume of lending does not explain the behavior of the provisions.

The "earnings before income tax, profit sharing, and loan loss provisions over total assets" variable was not found to be significant in Spain, at a 1% level of significance. The negative sign of the coefficient indicates growth of the provision whenever earnings decrease and vice-versa, thus going against expectations. There is no evidence of earnings management practices via provisions in the Spanish banks.

The "net equity over total assets" variable is not significant, thus going against the expectation with regards to the use of the provision for earnings management.

Contrary to the expectation of hypothesis H2, at a 1% level of significance the coefficient of the "GDP growth" variable was found to be significantly negative, indicating pro-cyclical behavior.

The "unemployment rate" variable was not found to be significant, going against expectations. The "size" control variable was found to be significant, with a positive sign, indicating that the larger the bank, the higher the provision tends to be.

### 6.3.3. H3: Banks that operate in Brazil.

The coefficient of determination ($R^2$) indicates that the independent variables of the model are associated with 15.65% of the dependent variable behavior (Table 14). The F statistic, which presents a p-value equal to 0.0000, confirms the statistical significance of the model.

**Table 14** *Coefficients of determination and F statistic of the model – Brazil.*

| Variables | Value |
|-----------|-------|
| $R^2$ | 0.1565 |
| $R^2$ Adjusted | 0.1531 |
| F Statistic | 45.1995 |
| P-Value (F) | 0.0000 |

**Source:** *Elaborated by the authors.*

According to the data in Table 15, the macroeconomic variables and the specific accounting variables of the banks studied in Brazil, with the exception of the unemployment rate and size, have significant effects on the provision.

**Table 15** *Regression results – Brazil.*

| Variable | Coefficient | Standard deviation | t-statistic | Prob |
|----------|-------------|--------------------|--------------|------|
| C | -0.0086 | 0.0080 | -1.0795 | 0.2805 |
| LOAN | 0.0268 | 0.0026 | 10.4186 | 0.0000 |
| ΔlnLOAN | -0.0014 | 0.0007 | -1.8981 | 0.0579 |
| EAR | 0.0868 | 0.0170 | 5.1125 | 0.0000 |
| NE | 0.0138 | 0.0045 | 3.0866 | 0.0021 |
| ΔGDP | -0.0477 | 0.0186 | -2.5637 | 0.0104 |
| UNEMP | -0.0173 | 0.0185 | -0.9356 | 0.3496 |
| SIZE | 0.0004 | 0.0004 | 0.9305 | 0.3523 |

**C** = constant in the regression; *LOAN* = balances of credit operations of banks i in time t; *ΔlnLOAN* = variation in the balances of credit operations of banks i in time t; *EAR* = earnings before income tax, profit sharing, and loan loss provisions over the average total assets of banks i in time t; *NE* = equity over the total assets of banks i in time t; *ΔGDP* = real growth of gross domestic product in time t; *UNEMP* = unemployment rate in time t; *SIZE* = size of banks i in time t.
**Source:** *Elaborated by the authors.*

The "loans over total assets" variable was found to be significant at a 1% level of significance. As expected, when a bank increases its share of credit operations in total investments, the percentage of the provision also increases, confirming the claim of growth in risks in phases of economic expansion. This prudent behavior of the banks contributes to attenuating the effects caused by possible pro-cyclical behaviors.

The "growth in lending" variable was only found to be significant at a 10% level of significance. The coefficient presented a negative sign, indicating that the provisions usually decrease when the balance of credit operations increases. This finding would go against the initial expectation, which indicated a probable positive relationship between growth in the lending volume and the provision.

The "earnings before income tax, profit sharing, and loan loss provisions over total assets" variable was also found to be significant at a 1% level of significance, indicating that variations in earnings are related with the behavior of the provision. The positive sign of the coefficient indicates growth in provision levels whenever earnings increase and vice-versa, thus supporting the expectation presented in Table 2. The result indicates the use of the provision for earnings management, a practice which ameliorates the pro-cyclical effects of the provision.

The "net equity over total assets" variable was shown to be significant at a 1% level of significance. The negative sign of the coefficient would go against the expectation of use of the provision for capital management; it is supposed that the banks raise the provision whenever their capital levels are more comfortable. It is assumed that the fact that the banks in Brazil have historically presented a higher reference equity than that required by the Central Bank and the absence itself of the generic provision instrument in the country's regulation explain this result.

Regarding hypothesis H3, at a 1% level of significance the coefficient of the "GDP growth" variable was shown to be significantly negative, indicating pro-cyclical behavior. This appears to indicate the absence of an efficient risk evaluation mechanism with a forward-looking character.

With regards to the "growth in the unemployment rate" variable, this was not found to be significant, thus going against the expectation presented in Table 2. The same occurs with the "size" control variable. It cannot be claimed that larger banks in Brazil, or those belonging to conglomerates, set more robust provisions than the rest.

As a sensitivity analysis, the lagged GDP and unemployment rate variables were added to the models. In Brazil and the United Kingdom, these variables were not found to be significant, indicating that the effect of their behavior on the provision calculation occurs in the same accounting period. With regards to Spain, the unemployment rate variable was not found to be significant, but the GDP variable was shown to be significant at a 1% level of significance.

## 7. FINAL REMARKS

Studies such as those by Bikker and Hu (2002), Laeven and Majnoni (2003), and Bikker and Metzemakers (2004) reveal that the choice between different accounting models would determine the behavior of the provision with relation to the business cycles. The expectation is that the incurred loss model leads to pro-cyclical behavior of the provision and the expected loss model results in countercyclical behavior of the provision. This study proposed to investigate whether the incurred loss models, which are conceptually pro-cyclical, would lead banks to setting pro-cyclical provisions, and whether the expected loss models, which are conceptually countercyclical, would lead banks to recognizing countercyclical provisions.

Two countries were intentionally chosen to represent these accounting models: Spain (expected loss) and the United Kingdom (incurred loss). A mixed model adopted in Brazil was also analyzed. The prevalence of typical procedures of expected loss models reinforced the hypothesis that the Brazilian model would be countercyclical.

Based on the linear regression models used in previous studies, especially those by Cavallo and Majnoni (2001), Bikker and Hu (2002), Laeven and Majnoni (2003), and Bikker and Metzemakers (2004), variables were selected to compose an econometric model that identified the relationship between the provision and the economic cycles in the three countries selected.

Two variables were inserted into the model in order to identify the relationship between the provision and economic cycles: the real variation in GDP and the unemployment rate. Contrary to expected, the unemployment rate was not found to be statistically significant in Brazil or in Spain. In United Kingdom this variable was found to be significant, at a 10% significance level, presenting a positive coefficient, which indicates that the provision tends to grow when the unemployment rate increases. It bears mentioning that studies such as those by Bikker and Metzemakers (2004) have already revealed that the unemployment rate would not be the best proxy to evaluate the relationship between the provision and economic cycles, a fact that is supported by this study.

The variable unanimously taken in the studies involving cyclicality is the variation in GDP. The findings of this research are consistent with these studies, as GDP was also found to be a statistically significant variable at a level of significance of practically 1% for each accounting model that was the object of this research.

With regards to the sign of the coefficient of the variable that reflects GDP, two of the formulated hypotheses were not supported. The biggest surprise was in Spain, whose accounting model prescribes setting forward-looking provisions, including provisions of a generic nature, which should cause countercyclical provisions. The sign of the coefficient of the variable, which would presumably be positive, was found to be negative, indicating pro-cyclicality. In Brazil, despite the general rule being to set forward-looking provisions, and although backward-looking previsions ultimately prevail when the delays in the payments of principal and interest are present, the study also revealed pro-cyclical behavior. It is inferred that the subjective character of the provision and the difficulty itself of predicting future economic scenarios, among other factors, could be leading to provisions of a different nature from that which the legislator intended when establishing more specific criteria for setting them. Only in relation to the United Kingdom was the hypothesis confirmed that the accounting model adopted, based essentially on the IASB rules, has in fact caused provisions that are pro-cyclical in nature.

Also in relation to the United Kingdom, only the GDP growth variable was found to be significant, at a 5% level of significance. It is assumed that the variables that represent credit risk, such as "loans over total assets" and "growth in lending" were not found to be significant due to the characteristics of the accounting model itself adopted in that country, which only recognizes the provision when objective evidence of loss is identified. Thus, the impact of the two variables would only occur indirectly (a greater share of loans over assets or growth in the volume of lending can lead to a higher probability of occurrence of objective evidence of loss in the future). These variables can, therefore, change without the banks in the United Kingdom having adjusted their provisions. In the United Kingdom, there was also no evidence found that the banks are using the provision for earnings management, thus contradicting the study by Silva (2016), or for capital management.

With regards to earnings management, this is assumed to be a reality for commercial banks in Brazil, thus supporting the results of the studies by Macedo and Kelly (2016), Bortoluzzo, Sheng, and Gomes (2016), and Silva (2016). The "earnings before income tax, profit sharing, and loan loss provisions" variable was found to be significant, at a 1% level of significance, having presented a positive coefficient indicating that the provision tends to grow when the banks' earnings increase. In relation to Spain, despite the significance at a 1% level of significance, this variable presented a negative sign, signaling that the provision decreases when earnings increase and thus contradicting the study by Silva (2016).

Earnings management practices, which are reprehensible in some aspects, can represent a counterpoint to setting pro-cyclical provisions, easing their effects on the current phase of the business cycle.

Capital management practices were also not found to be a reality for the commercial banks in Brazil and Spain. In Brazil, despite the significance at a level of 1%, the aforementioned variable presented a positive coefficient, indicating that the greater the share of net equity over total assets, the greater the provision set by commercial banks. This situation may be indicating that the comfortable level of capitalization presented by the banks in Brazil (23.20% of total assets, on average) is inducing an increase in the provision level whenever net equity presents a positive variation, which represents a conservative procedure that also serves as a counterpoint to setting pro-cyclical provisions. In Spain, the "net equity over total assets" variable was not found to be significant, indicating that the provision level is not affected by the behavior of the equity of commercial banks that operate in that country.

The "size" variable was found to be statistically significant only in relation to the banks that operate in Spain. In the other countries, the size of the banks does not appear to influence the provision level.

The results support the previous studies by Bikker and Hu (2002), Laeven and Majnoni (2003), and Bikker and Metzemakers (2004), in the sense that the provision behavior strongly depends on the economic cycle, indicating that the provision generally grows in poor economic times. Another similarity between the results of this research and the aforementioned studies is the finding that the effect of pro-cyclicality is usually mitigated by an increase in the provision in periods of higher earnings (earnings management practices).

A possible assumption regarding the fact that the research revealed pro-cyclical behavior when countercyclical behavior was expected is that the banks may be using discretion and managing their earnings via the provision and more than compensating for the expected effect of the models, which is a hypothesis to be investigated in future studies.

It was also found that in Brazil and the United Kingdom the effect of economic variations on the provision behavior takes place in the same accounting period. In the case of Spain, however, a variation in GDP in the previous period was found to be a significant variable.

With relation to the econometric model chosen for the research, it was found that the peculiar nature of the provision rules in the United Kingdom, in relation to those in Brazil and Spain, may be determining the difference in the levels of significance of the explanatory variables for the model, when this is applied to commercial banks in the country that adopts the IASB accounting standards. Thus, it is assumed that adaptations are needed to adjust the model used to the specificities of countries that use relatively different accounting rules for setting the provision.

As an intentional sample is concerned, which considers only one country for each accounting model, the results may not be generalized. For future research, studies with different econometric models are suggested, incorporating variables not considered in this research, especially those related to the delay in the payment of principal and interest and to the accounting recognition of losses. Evaluating the question of cyclicality by considering different accounting scenarios in the same country (for example, studying cyclicality before and after the advent of Resolution n. 2,682/1999) could also prove to be an interesting experiment for reflecting on the provision behavior in different regulatory environments.